\documentclass[12pt]{article}
\usepackage{a4wide,amssymb}

\newcommand{\be}{\begin{equation}}
\newcommand{\ee}{\end{equation}}
\newcommand{\eref}[1]{(\ref{#1})}

\newcommand{\ms}{m_\mathrm{s}}
\newcommand{\mv}{m_\mathrm{v}}
\newcommand{\mpl}{M_\mathrm{Pl}}
\newcommand{\GeV}{\,\mathrm{GeV}}
\newcommand{\mm}{\, \mathrm{mm}}
\newcommand{\Or}{\mathcal{O}}

\begin{document}

\title{\bf Brane World Cosmic String Interaction}

\author{Stephen C. Davis\footnote{sdavis@lorentz.leidenuniv.nl} \\
\small \em  Instituut-Lorentz for Theoretical Physics, Postbus 9506,\\
\small \em NL--2300 RA Leiden, The Netherlands.
}

\maketitle

\begin{abstract}
It is shown that brane world gravity produces an attractive force between
cosmic strings, in contrast to conventional Einstein
gravity. This force is sufficient for stable bound states of
Type II strings (which normally repel each other) to form. There is a
small separation between the strings, giving them a rope-like structure.
At astronomical scales these `cosmic ropes' resemble higher
winding number strings, and so all brane world cosmic strings
are effectively Type I.
\end{abstract}

There has been a renewed interest in the line-like topological defects
known as cosmic strings~\cite{book}, particularly since they have been found to
occur in many superstring theory scenarios~\cite{strstr}. They may
also be making a small contribution to the cosmic microwave background (CMB)
radiation. Cosmic strings have been ruled out as the dominant source of
CMB anisotropies, but their contribution could still be large enough to
be detected by upcoming experiments.

Most string network simulations, from which the above result is
obtained, assume that at short distances two cosmic strings
repel. This means that only strings with winding number $n=1$ are
stable, and that when two strings collide they will intercommute
(exchange partners). However this is not true for all
strings. Denoting the masses of the string's Higgs and gauge bosons by
$\ms$ and $\mv$, the above statement only holds for Type II strings,
for which the ratio $\beta = \ms^2/\mv^2 > 1$. On the other hand Type
I strings, for which $\beta <1$, are attracted to each other. Higher
winding number cosmic strings are stable in this case, and colliding
strings do not always intercommute~\cite{book,tom}.

When considering the force between two strings, the effects of gravity
are generally ignored. For conventional Einstein gravity this is
completely reasonable, as in this case the force is
zero. However this need not be true for more exotic gravity
theories. Consider the superstring motivated brane world scenario, in
which our apparently 3+1 dimensional universe is a 3-brane embedded in
a higher dimensional, bulk spacetime. The Randall-Sundrum
scenario~\cite{RSII}, which has a single warped extra dimension,
provides the simplest example of this. In this case the extra
dimensional effects alter Newton's law at short distances. In
ref.~\cite{BWCS} it was shown that they also produce an attractive
force between cosmic strings and ordinary matter.

In this letter it will be shown that cosmic strings on brane worlds
attract each other gravitationally. Close to the string core this
force is typically weaker than the usual string forces, although its range is
greater. Type II strings will therefore be attracted by each
other. Since at very short distances they still repel, groups of parallel
strings will form into bound states with small gaps between the different
strings, producing a kind of `cosmic rope'. At larger distances these
configurations will resemble a single string with winding number $n>1$.

Note that the brane world cosmic strings in this article are assumed to be
formed of fields which are completely confined to the brane. Cosmic
string solutions which extend into the bulk space can also be
constructed, and will also have novel gravitational
properties. However they will not considered here.

The equations of motion for a cosmic string are
\be
\mu \left[ \frac{1}{\sqrt{-\gamma}}
\left(\sqrt{-\gamma} \gamma^{ab} x^\nu_{,b}\right)_{,a} 
+ \Gamma^\nu_{\rho\sigma} \gamma^{ab}  x^\rho_{,a} x^\sigma_{,b}\right] = F^\nu
\, .
\label{streq}
\ee
$\mu$ is the energy per unit length of the string, and is of order $\ms^2$
for Type II strings. $F^\nu$ is any non-gravitational force
acting on the string, and
\be
\gamma_{ab} = g_{\mu\nu}x^\mu_{,a} x^\nu_{,b}
\ee
is the string worldsheet metric. The index $a$ runs over the two
worldsheet coordinates $\zeta^0$, $\zeta^1$.
We will be considering the motion of one string around second parallel
string, and work in the rest frame of the second string. In the
absence of an inter-string force and gravitational effects, we could
choose $\zeta^0=t$ and $\zeta^1=z$.

The force between the two strings is $F^\nu
= - \partial^\nu E_\mathrm{int}(r)$, where $E_\mathrm{int}$ is the
interaction energy of two strings. This energy is respectively
positive and negative for $\beta>1$ and $\beta <1$. For strings with
strong Type II behaviour ($\beta \gg 1$), the approximate
expression~\cite{book,int}
\be
E_\mathrm{int} \approx \mu K_0( \mv r) \sim
\mu \left(\frac{\pi}{2\mv r}\right)^{1/2} e^{-\mv r}
\label{Eint}
\ee
holds away from the string core, whose size is of order $\mv^{-1}$.

In ref.~\cite{BWCS} it was shown to leading order in the
gravitational coupling, that (on the brane) the metric around a cosmic
string is
\be
ds^2 = (-dt^2+dz^2)[1- G\mu f_2(r/\ell)] + dr^2 
+ r^2(1-8G\mu[1+f_1(r/\ell)]) d\theta^2 \, .
\label{metric}
\ee
In conventional Einstein gravity the functions $f_1$ and $f_2$ would
be identically zero.

The length scale $\ell$ is related to the warp factor of the bulk. It
is related to the underlying 5-dimensional gravity scale $M_5$ and the
effective four dimensional Planck mass $\mpl$ by $\ell = \mpl^2/M_5^3$. 
It also gives length scale below which deviations from conventional
gravity will be significant. Table-top test of Newton's law constrain
it to $\ell \lesssim 0.1 \mm$, which implies $M_5 \gtrsim 10^8 \GeV$.

For a grand unified theory (GUT) cosmic string, $\ms \sim 10^{16}
\GeV$. However it only makes sense to consider such a string in the
above brane scenario if $\ms < M_5 < \mpl$, since otherwise the low
energy effective gravity theory we are using will not be valid near
the string. Hence for our theory to have GUT strings we must have $M_5
\gtrsim 10^{16} \GeV$ or $\ell \lesssim 10^{-23} \mm$.

For strings with $\ell \mv > 1$, the above functions have the
approximate form
\be
f_1 =  -\frac{2\ell}{3r}\ln\frac{\ell}{r} + \Or(1)
\, , \qquad
f_2 =  \frac{4\ell}{3r} + \Or\left(\ln\frac{\ell}{r}\right)
\label{appfa}
\ee
when $\mv^{-1} \ll r \ll \ell$. 
At larger distances ($r \gg \ell$) the asymptotic behaviour of the
functions is 
\be
f_1 =  -\frac{2\ell^2}{3r^2}
+\Or\left(\frac{\ell^4}{r^4} \ln\frac{\ell}{r}\right)
\, , \qquad
f_2 =  \frac{4\ell^2}{3r^2}
+\Or\left(\frac{\ell^4}{r^4} \ln\frac{\ell}{r}\right) \ . 
\label{appfb}
\ee
For lighter strings ($\mv \ell <1$), the
expressions~\eref{appfb} are valid everywhere outside the string.
We see that as $r\to \infty$, the metric~\eref{metric} reduces to the
usual conical cosmic string metric.

We will now find approximate solutions of \eref{streq} when
gravitation and the string interaction are included. We are interested
in a system of two parallel strings, so we take $d r/d\zeta^1 =
d\theta/d\zeta^1=0$. We see that
the $t,z$ components of \eref{streq} are satisfied by
$z=\zeta^1 + \Or(G\mu)$, $t=\zeta^0+ \Or(G\mu)$. The
remaining components then imply
\be
r^2 \dot \theta \approx J [1 + \Or(G\mu)]
\ee
\be
\ddot r \approx G\mu \partial_r f_2 + \frac{J^2}{r^3}
- \frac{\partial_r E_\mathrm{int}}{\mu}+ \Or(G\mu)^2
\ee
for some constant $J$. The inter-string distance $r$ is assumed to be
in the region where  $E_\mathrm{int}/\mu \sim G \mu$. For a Type II
string we find that the above equation has solutions with $r$ constant. 

Using the approximate expression for $f_2$~\eref{appfa}, the distance between
the strings in this case satisfies
\be
(\mv r)^{3/2} e^{-\mv r} = \sqrt{\frac{2}{\pi}}
\left(\frac{4 G \mu \ell \mv}{3} - \frac{J^2\mv}{r}\right)
\label{rora}
\ee
For a static configuration $J=0$. In this case \eref{rora} has
two solutions if the right hand side is less than
$[3/(2e)]^{3/2}$. Since $G \mu \ell \mv \lesssim \ms^3/M_5^3 <1$, this
is generally the case. In fact this quantity will usually be very
small, in which case the stable, larger $r$ solution of \eref{rora} is
approximately 
\be
r \approx -\frac{1}{\mv} \ln\frac{16 G \mu \ell \mv}{9 \sqrt{3\pi}} \, .
\ee
The separation of the strings is then a few times larger
than the string core. The other solution is generally
inside the string core, where the approximations used to derive
\eref{rora} are not valid. Even if it is outside the core (which can only
happen if $G \mu \ell \mv \sim 1$), it is unstable.

If on the other hand the string core is larger than $\ell$, which is
the case if the string is light or the five and four dimensional
Planck scales are close, then the asymptotic expression~\eref{appfb}
for $f_2$ will apply. Constant $r$ solutions will then satisfy
\be
(m_v r)^{5/2} e^{-\mv r} = 
\sqrt{\frac{2}{\pi}} \left(\frac{8 G \mu \ell^2\mv^2}{3} - J^2\mv^2\right) 
\, .
\label{rorb}
\ee
Since $\ell \mv <1$, the right hand side of this will
generally be small, and the equation will have two solutions. For
$J=0$ there is a stable one at
 \be
r \approx -\frac{1}{\mv} \ln\frac{64 G \mu \ell^2 \mv^2}{75\sqrt{5\pi}} \, ,
\ee
and an unstable one which is usually has $\mv r <1$, and so is not
relevant.

If $G \mu \ell\mv$ is too large there will be no constant $r$
solution. This will only be the case for very heavy strings. The
attractive gravitational force is the dominant effect in this case, and
so the strings will act like Type I strings, even when 
$\beta$ indicates they are Type II. Even for lighter string bound
states with a non-zero separation, the pair of strings will resemble an
$n=2$ string when viewed at large (e.g.\ astronomical) scales, so these
strings are effectively Type I as well.

As well as the above static configurations, the two strings could also
be orbiting each other (i.e.\ $ J>0$). Starting with \eref{rora}, we
see that as $J$ increases the radius of the orbit increases, until $r >\ell$ 
(when $J^2 \sim G \mu \ell^2$). At this point \eref{rora} is no longer
valid and \eref{rorb} applies instead. From \eref{rorb}, we again see that
the radius of the orbit increases with $J$, up until 
$J_{\rm max} =(8 G \mu \ell^2/3)^{1/2}$. Faster moving strings will
not form bound states.

Taking $J$ to be non-zero also alters the small $r$ behaviour of
\eref{rora}. The unstable solution may disappear. An additional stable
solution may also appear. However all of this occurs inside, or almost
inside the string core. Since the approximations we are using are not
valid there, these results should not be trusted.

We have seen that slowly moving Type II cosmic strings will be
attracted to each other by brane world gravity. They will then form
into a sort of cosmic rope, with the separation of the strings being a
multiple of the string radius. This stable configuration could also carry a
small amount of angular momentum. At larger distances this rope will
resemble a higher winding number cosmic string, and so on astronomical
scales, Type II strings act as though they are Type I. For Type I and
critical ($\beta=1$) strings, the only inter-string forces are
attractive and so the usual Type I results apply, even with the
gravity modifications. Hence brane world cosmic strings are all
effectively Type I.

This appears to suggest that there is a serious conflict between the
existence of cosmic strings and the brane world scenario. If higher
winding number cosmic strings are stable, it is not certain that
colliding cosmic strings will intercommute. Without intercommuting an
evolving string network will not be able to loose energy by loop
production. This suggests it will not reach a scaling solution, and
will end up dominating the energy density of the universe, contrary to
what is observed.

In fact this potential problem is not so serious. Firstly, it has been
shown that it is possible for non-intercommuting strings to reach a
scaling solution~\cite{DF}.  Furthermore, a careful study of Type I
strings reveals that in fact they do intercommute, except when the
angle between the colliding strings is small~\cite{tom,dani}. Hence if
our universe is a brane world, the presence of a cosmic string network
on it need not conflict with any current observational constraints.

In the cases when strings do not intercommute, the collision
will instead result in an `zipper' configuration, consisting of a
segment of $n=2$ string which separates into two $n=1$ strings at each
end. The resulting string network will therefore contain Y-junctions.
Observation of such junctions could  provide indirect evidence for the
brane world scenario. Of course this assumes that other explanations
for the Y-junction, such as the strings simply being Type I, have been
excluded by other observations or theoretical arguments. An additional
property of brane world cosmic stings is that ordinary matter is
attracted to the strings~\cite{BWCS}. This does not occur for
conventional strings (Type I or otherwise), and may provide a way to
distinguish between string junctions which arise from brane effects,
and junctions with other explanations.

\ \\

I am grateful to Paul Shellard for useful comments and to the
Netherlands Organisation for Scientific Research (NWO) for financial support.

\end{document}